\documentclass[11pt,twoside]{article}  

\usepackage{asp2014}

\aspSuppressVolSlug
\resetcounters

\begin{document}

\title{From Photometric Redshifts to Improved Weather Forecasts: machine learning and proper scoring rules as a basis for interdisciplinary work}

\author{Kai~Polsterer,$^1$ Antonio~D'Isanto,$^1$ and Sebastian~Lerch$^{1,2}$}
\affil{$^1$HITS gGmbH, Heidelberg, Germany; \email{kai.polsterer@h-its.org}}
\affil{$^2$KIT, Karlsruhe, Germany}

\paperauthor{Kai~Polsterer}{kai.polsterer@h-its.org}{0000-0002-3435-1912}{HITS gGmbH}{AIN}{Heidelberg}{BW}{69198}{Germany}
\paperauthor{Antonio~D'Isanto}{antonio.disanto@h-its.org}{0000-0002-7093-439X}{HITS gGmbH}{AIN}{Heidelberg}{BW}{69198}{Germany}
\paperauthor{Sebastian~Lerch}{sebastian.lerch@h-its.org}{0000-0002-3467-4375}{HITS gGmbH}{CST}{Heidelberg}{BW}{69198}{Germany}


\begin{abstract}
    The amount, size, and complexity of astronomical data-sets and databases are growing rapidly in the last decades, due to new technologies and dedicated survey telescopes.
    Besides dealing with poly-structured and complex data, sparse data has become a field of growing scientific interest.
    A specific field of Astroinformatics research is the estimation of redshifts of extra-galactic sources by using sparse photometric observations.
    Many techniques have been developed to produce those estimates with increasing precision.
    In recent years, models have been favoured which instead of providing a point estimate only, are able to generate probabilistic density functions (PDFs) in order to characterize and quantify the uncertainties of their estimates.
    
    Crucial to the development of those models is a proper, mathematically principled way to evaluate and characterize their performances, based on scoring functions as well as on tools for assessing calibration.
    Still, in literature inappropriate methods are being used to express the quality of the estimates that are often not sufficient and can potentially generate misleading interpretations.
    In this work we summarize how to correctly evaluate errors and forecast quality when dealing with PDFs.
    We describe the use of the log-likelihood, the continuous ranked probability score (CRPS) and the probability integral transform (PIT) to characterize the calibration as well as the sharpness of predicted PDFs.
    We present what we achieved when using proper scoring rules to train deep neural networks as well as to evaluate the model estimates and how this work led from well calibrated redshift estimates to improvements in probabilistic weather forecasting.
    The presented work is an example of interdisciplinarity in data-science and illustrates how methods can help to bridge gaps between different fields of application.
\end{abstract}



\section{Introduction}

By applying techniques from the fields of computer sciences, mathematics, and statistics, astronomical data can be accessed and analyzed more efficiently.
A specific field of research in Astroinformatics is the estimation of the redshift $z$ of extra-galactic sources, a measure of their distance, by just using sparse photometric observations \citep{2019arXiv190407248B}.
For many scientific applications, especially those from cosmology, photometrically derived redshifts are crucial \citep{2011arXiv1110.3193L, 2014MNRAS.445.1482S}.
To measure $z$ directly, the spectral information of a source has to be collected.
Therefore the light has to be dispersed, making the observation a technically and exposure time wise demanding task.
Spectroscopically obtained redshifts are available for only a limited number of sources, making those values valuable to train and evaluate statistical models.
In recent times, those models became more and more common to derive $z$ from photometric measurements, especially for larger samples where a direct observation is still not feasible.
Many techniques have been developed to produce those estimates with increasing precision \citep{2019NatAs...3..212S}.

In the past, empirical, on training data based machine learning approaches, usually retrieved $z$ as a point estimate not reflecting any uncertainties.
On the other hand, template fitting based methods \citep{2000A&A...363..476B} which where reporting a likelihood were often lacking the predictive performances of the empirical ones.
In the field of empirical models, instead of providing a point estimate only, astronomers started to generate uncertain estimates \citep{2013MNRAS.432.1483C, 2016PASP..128j4502S}.
Because the integration of the spectral information through a set of photometric filters leads to an irreversible, degenerate problem, often a more complex description of the estimates as full probabilistic density functions (PDFs) is mandatory.
In contrast to dealing with just a single prediction, this enables scientists to evaluate the likelihood for different redshifts.
Building accurate statistical models is a mandatory step, especially when it comes to reflecting the uncertainty of the estimates.
This demands a correct characterization of the performance of the predictive model and the use of adequate evaluation tools.

The evaluation of predictive models is a common and important aspect across scientific disciplines, and the proliferation of probabilistic forecasting methods has led to a wide variety of available validation methods.
In the field of weather forecasting, besides the log-likelihood, the continuous ranked probability score (CRPS) as well as the probability integral transform (PIT) are common tools used for evaluation.
Research on mathematical properties of proper scoring rules and related verification metrics has become an active area of statistical research \citep{gneiting2007probabilistic}.
In \citet{2018A&A...609A.111D} proper evaluation tools have been introduced to astronomy, which allow a fair comparison of the achieved prediction quality.
The proposed tools have been utilized to develop a model that self calibrates and produces reliable PDFs as outcome \citep{DBLP:conf/esann/DIsantoP17}.
In turn, this idea was adapted in weather forecasting to develop a new approach to ensemble post-processing \citep{epub67034}.
This is an excellent example how methodology can help to bridge the gap between completely different sciences.
The interdisciplinary dialogue which led to both publications suggests that the implementation of these methods in other fields beyond astronomy and weather forecasting has the potential to improve probabilistic predictions.

Despite the mathematically principled tools available for forecast evaluation, many publications in the field of photometric redshift estimation use inappropriate and custom tools and measures that do not fulfill essential criteria of proper scoring rules.
Often tools tailored to point estimates have been extended to the usage on PDFs without considering the involved implications.
Therefore this work gives a detailed overview of proper evaluation and discusses common mistakes and misconceptions.
Such a discussion is fundamental, as the success of many forthcoming project and missions will be highly dependent on the availability of reliable and affordable redshift measurements.

The publication is structured as follows:
After the introduction in Sec.~\ref{evaluation} proper evaluation tools are presented and discussed.
In Sec.~\ref{evaluation} we discuss common mistakes which can be found in many publications.
The neural network we developed for photometric redshift estimation is shown in Sec.~\ref{DCMDN} which lead to a new approach of post-processing of weather forecasts which is summarized in Sec.~\ref{weather}.
Sec.~\ref{conclusion} concludes this work.

\section{Proper Evaluation}
\label{evaluation}

The evaluation of predictions always depends on the availability of observed true values.
In the case of photometric redshift estimation this corresponds to the spectroscopically determined redshift values.
Usually one can ignore the measurement errors of those redshifts, even though the automatic pipelines might introduce errors.
Instead of considering the measured redshift as a PDF itself, we treat it as a deterministic realization.

\subsection{Calibration, Sharpness and Proper Scoring Rules}\label{proper}

When it comes to a correct evaluation of the performance of probabilistic predictions, it is mandatory to fulfill the paradigm of “maximizing the sharpness of the predictive distributions subject to calibration” \citep{gneiting2007probabilistic}.
Calibration refers to the statistical consistency between a predicted distribution and the observed true value, and formalizes the idea that the observed value should be indistinguishable from a random draw from the forecast distribution.
Sharpness expresses how much the same predicted distribution is concentrated.
Calibration can be seen as a joint property between predictions and true values, while sharpness is a property of the predictions only.
The more concentrated the predictions are, the sharper they are.
Hereby it is important that the sharpness always has to be considered together with the calibration.
These two attributes are important indicators of the overall quality of a prediction.

A scoring rule assigns a numerical value (a score) to a pair of a probabilistic forecast and an observation, and is called proper if in expectation, the true distribution of the observation receives the best possible score.
The definition can be formalized \citep{RePEc:bes:jnlasa:v:102:y:2007:p:359-378}  by stating that

\begin{equation}
s(G, G) \geq s(F, G),
\end{equation}

\noindent reflecting that the score $s$ between the natural distribution $G$ and itself always has to be larger or equal to any forecast distribution $F$. 
The ideal and best score is reached when $F=G$ assuming that $s$ is better the larger it is.
If this maximum is unique, then the scoring rule is called strictly proper.
Proper scoring rules are important tools for evaluating probabilistic forecast as they can address calibration and sharpness simultaneously.

\subsection{Log-Likelihood}
\label{likelihood}

\begin{figure}
  \includegraphics[width=\columnwidth]{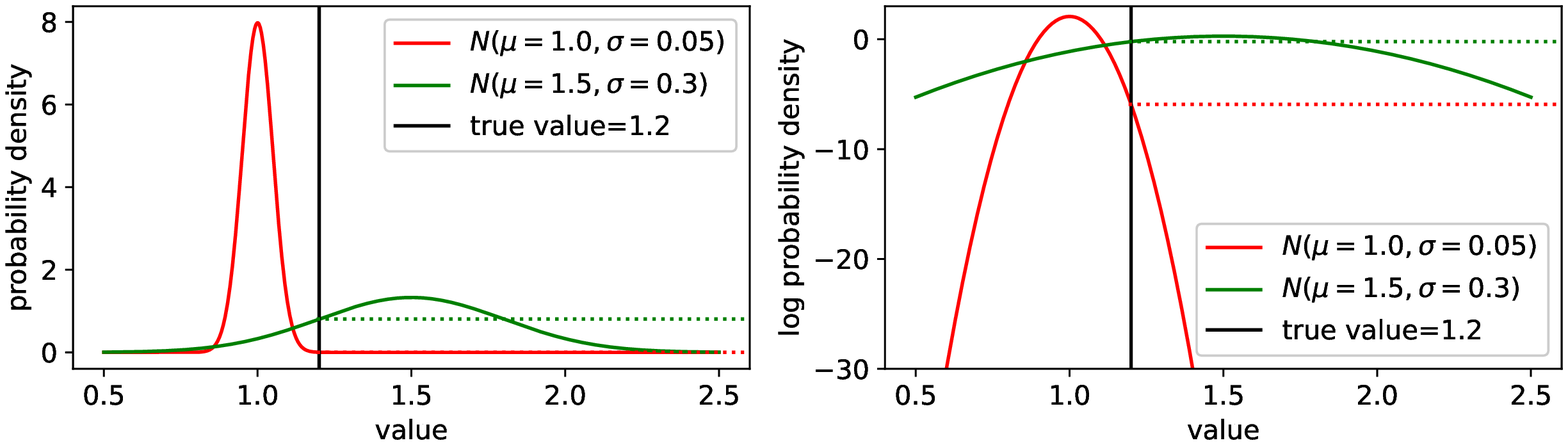}
    \caption{Visualization of the likelihood (left) and the log-likelihood (right) for two different normal distributions evaluated at a given true value.}
    \label{fig:likelihood}
\end{figure}

The concept of likelihood was introduced by Fischer as a basis to deal with inference problems.
When evaluating the estimated PDF for a given source, it helps to understand how likely the estimate was given the true redshift value $z$.
It is clear, that the higher the likelihood of the estimate at the true value is, the better the estimate is.
Usually this concept is used to optimize predictive models by minimizing the negative natural logarithm of the likelihood.
Obviously an estimate that is just slightly off the actual value but shows an over confident (i.e., too low) uncertainty will have a very low likelihood while at the same time a more uncertain estimate will have a higher one.
Note that the distribution in Fig.~\ref{fig:likelihood} with a higher bias in its mean value still exhibits a higher likelihood at the true value.
The log-likelihood as a scoring function is just improving when the predicted distribution is approaching the natural distribution.
Therefore the log-likelihood is a proper scoring rule.

\subsection{Continuous Ranked Probability Score}
\label{CRPS}

\begin{figure*}
  \includegraphics[width=\textwidth]{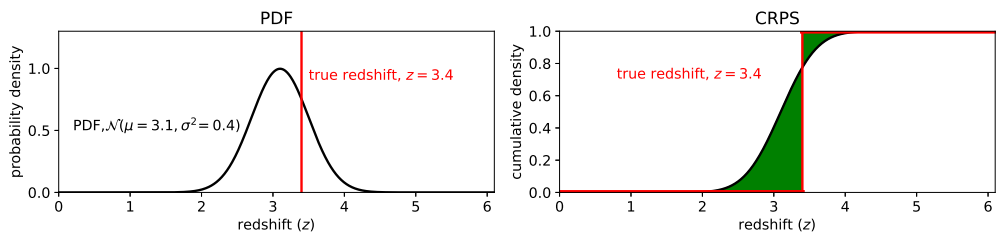}
    \caption{Visual guide to the relation between probability density function (PDF), cumulative distribution function (CDF), and continuous ranked probability score (CRPS).
    The true redshift that is used for calculating the CRPS, is plotted as a reference in red.
    The integral over and under the (CDF) that is used for the calculation of the CRPS is coloured in green.
    Note the Heaviside step-function, the CDF of the true redshift is marked red in the CRPS plot (right).
    }
    \label{fig:crps}
\end{figure*}

The CRPS \citep{Hersbach} is widely used in environmental sciences for expressing a discrepancy between a predicted PDF and a true value.
It compares a full distribution with an observation as defined by:

\begin{equation}\begin{split}
\text{CRPS} = \frac{1}{N}\sum_{t=1}^{N}\text{crps}(\text{CDF}_{t}, z_{t}), \\
\text{with } \text{crps}(\text{CDF}_{t}, z_{t}) = - \int_{- \infty}^{+ \infty}\left[\text{CDF}_{t}(z) - \text{CDF}_{z_{t}}(z)\right]^{2}dz
\label{eqn:CRPS}
\end{split}
\end{equation}

\noindent
$\text{CDF}_{t}$ is the cumulative distribution of the $\text{PDF}_{t}$, as defined in:

\begin{equation}
\text{CDF}_{t}(z_t) = \int_{- \infty}^{z_{t}}\text{PDF}_t(z)dz
\label{eqn:CDF}
\end{equation}

\noindent
In \mbox {Eq.~\ref{eqn:CDFz}} the cumulative distribution of the true redshift $\text{CDF}_{z_{t}}$ is defined based on $H(z) = \mathcal{H}$, the Heaviside step-function.
Note that here the CRPS is defined in negative direction, meaning that larger values are better.

\begin{equation}
\text{CDF}_{z_{t}}(z) = H(z - z_{t}), \text{ with } H(z) = \left\{
    \begin{array}{lr}
      0 & \text{for } z < 0 \\
      1 & \text{for } z \geq 0
    \end{array}
    \right. 
\label{eqn:CDFz}
\end{equation}

\noindent
The calculation of the CRPS as well as a Gaussian PDF and the corresponding CDF are visualised in \mbox{Fig.~\ref{fig:crps}}.
In case the PDFs are given as normal distributions, we are able to calculate it in a closed form \citep{Gneiting2005}.
The details of calculating the CRPS of a Gaussian mixture model in a closed form are given in \citet{QJ:QJ2006132621104}.
A comprehensive collection of analytical expressions of the CRPS for a wide variety of parametric families of probability distributions is available in \citet{JordanEtAl2019}.

\subsection{Probability Integral Transform}
\label{pit}

\begin{figure*}
  \includegraphics[width=\columnwidth]{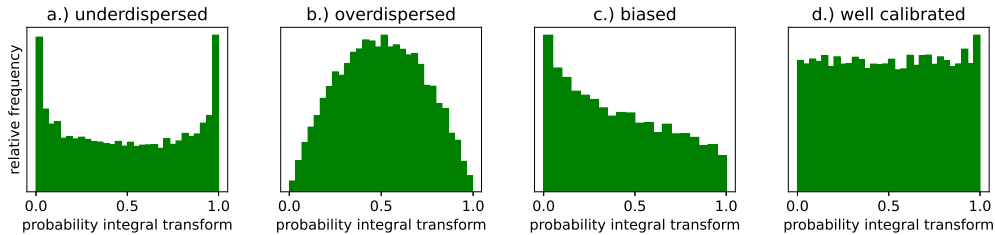}
    \caption{Four different histograms of PITs.
    In the case of underdispersed predictions an u-shaped, concave distribution is observed (a).
    Overdispersed predictions result in a peaked, convex distribution (b).
    When a slope in the histogram is observed, the analyzed predictions are biased (c).
    Only when the histogram exhibits a flat distribution, the PDFs are well calibrated (d).
    }
    \label{fig:pit}
\end{figure*}

As stated by \citet{gneiting2007probabilistic}, when optimizing the forecast performance of a model, the goal is to maximize the sharpness of the predictive distributions subject to calibration.
While sharpness is expressing the concentration of a probabilistic prediction, the calibration captures many aspects of the forecast with respect to the true values.
In the context of photometric redshift estimation this refers to the consistency between the predictive distribution and the true redshift.

In \citet{dawid1984present} the probability integral transform (PIT) was proposed to be used as a diagnostic tool to check the calibration of the generated predictive distributions.
It can be used in a visual tool which is based on the histogram of the values of the cumulative probability at the true value (see Eq.~\ref{eqn:CDF}).
With respect to photometric redshift estimation, the PIT is calculated with the CDF of the predicted distribution evaluated at the true redshift $z_{true}$.
A single PIT value is only carrying the information about the probability of the estimate to be smaller than or equal to the single true value which itself does not help to quantify the predictive performance as a score.
However, by inspecting this property for multiple instances, systematic effects can be examined.
When plotting the PIT values of all predictions of a model as a histogram, a diagnostic tool to visually check the calibration is provided (see Fig.~\ref{fig:pit}).
In case the predictions are well calibrated, the distribution of all CDF$(z_{true})$ values has to be uniform.
This visual check of calibration is based on the mathematical result that if a random variable $Y$ follows a probability distribution with CDF $G$, then the random variable $G(Y)$ follows a standard uniform distribution.
In case of underdispersion, overdispersion or biases of the estimates, characteristic shapes can be identified.

\section{Common Evaluation Mistakes}

The estimation of $z$ based on photometric features has a long history which started with point estimates.
Driven by the scientific goals, astronomers wanted to achieve, existing scores have been utilized or new have been formulated in order to compare different predictive models.
As more and more models were developed that produced probabilistic forecasts, some of the scores initially intended for point estimates have been extended to measure their performance.
In literature we find common conceptual mistakes with respect to evaluation of PDFs.

\subsection{Traditional Scores and Oversimplification}

The quality of photometric redshift estimates had been traditionally evaluated by adopting a certain number of statistical measures which evaluate the error between the true value and the prediction.
There are no particular problems in measuring the deviation between the true value and an estimated value and these scores are perfectly adapt when the prediction is based on point estimates.
The most commonly used ones are the bias, the root mean square error (RMSE), the median absolute deviation (MAD) and the standard deviation (STD) of the estimation errors.

A PDF captures the whole uncertainty of an estimate as it allows to calculate the probability for any given value range.
In astronomy, we \emph{must} be able to ask a system: "give me all sources with a redshift larger than $z=3.4$ with a probability larger than 90\%".
To be comparable to older publications, often the performance on point estimates is calculated by generating point estimates from the PDF.
This is done by calculating the mean, selecting just a single mode or doing other arithmetic operations.
Measures of discrepancy between the derived point estimate and the true value such as the RMSE are then used for evaluation.
While this technically can constitute a proper scoring rule if the computed functional and employed evaluation metric are carefully matched \citep{Gneiting2011}, this oversimplification does not provide an informed evaluation of the PDF and the inherent forecast uncertainty.

\subsection{Handling of Outliers}

\begin{figure}
  \includegraphics[width=\columnwidth]{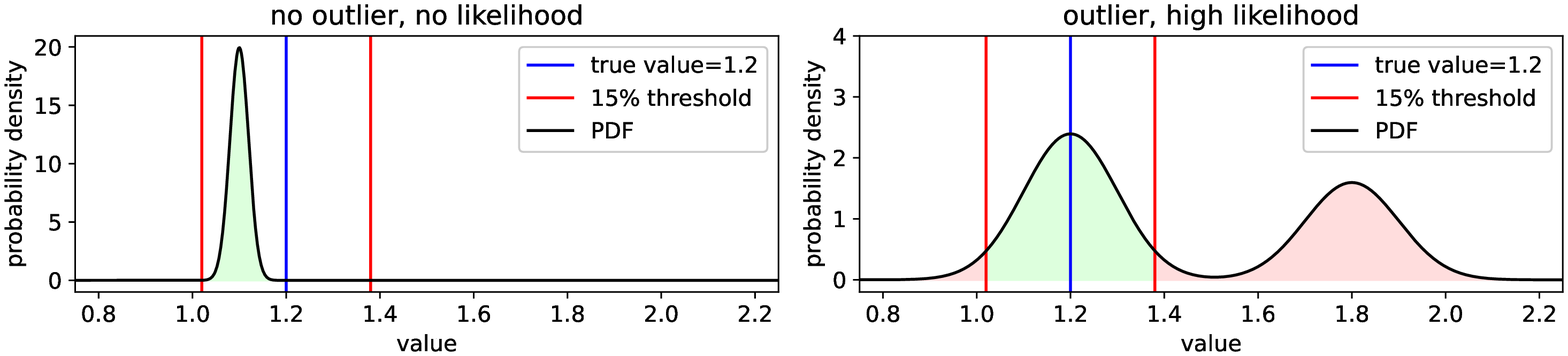}
    \caption{Example of \emph{wrong outliers treatment}.
    (left) Normal distribution with $\mu = 1.1, \sigma = 0.02$.
    Even though it is in the margins of $\pm 15\%$, this estimate has a likelihood of nearly zero at the true redshift $z = 1.2$.
    (right) GMM with $\mu_{1}=1.2, \sigma_{1}=0.1, \omega_{1}=0.6, \mu_{2}=1.8, \sigma_{2}=0.1, \omega_{2}=0.4$.
    Note the high likelihood, but due to the multimodality of the prediction a large area is out of the margins.
    }
    \label{fig:outliers}
\end{figure}

Many publications in the past used the distance between the point estimates and the true values as an indicator to find so called catastrophic outliers.
Estimates that had relative errors $\Delta z/(1+z)$ with $\Delta z=||z-z_{true}||$ larger than $15\%$, $5\%$ respectively, had been considered unreliable with respect to several cosmological applications.
When transferring this concept to probabilistic estimates, astronomers started aligning and coadding the estimated PDFs and defining outliers based on the integrated probabilities out of a defined threshold margin of $15\%$, $5\%$ respectively.

In a probabilistic setting, however, this definition of outliers does not hold.
Especially in the case of photometric redshift estimation, the reconstruction of the complex spectral information based on integrated broad band filters is not uniquely possible, making a complex, multimodal description of the probabilistic prediction necessary.
In cases where a distinct assignment of a single likely redshift is simply not possible, one mode of the PDF could match perfectly, while other modes would contribute to the thresholded outliers area (compare Fig.~\ref{fig:outliers}).

A better approach to deal with sources that do not show a single sharp prediction, would be either to use the generated PDFs to exclude those sources from further analysis but risking to add selection biases, or to further improve the predictive model by e.g.\ adding more photometric features.
The traditional concept of outliers \emph{cannot} be directly transferred to probabilistic estimates.

\subsection{Use of Improper Scores}

\begin{figure}
  \includegraphics[width=\columnwidth]{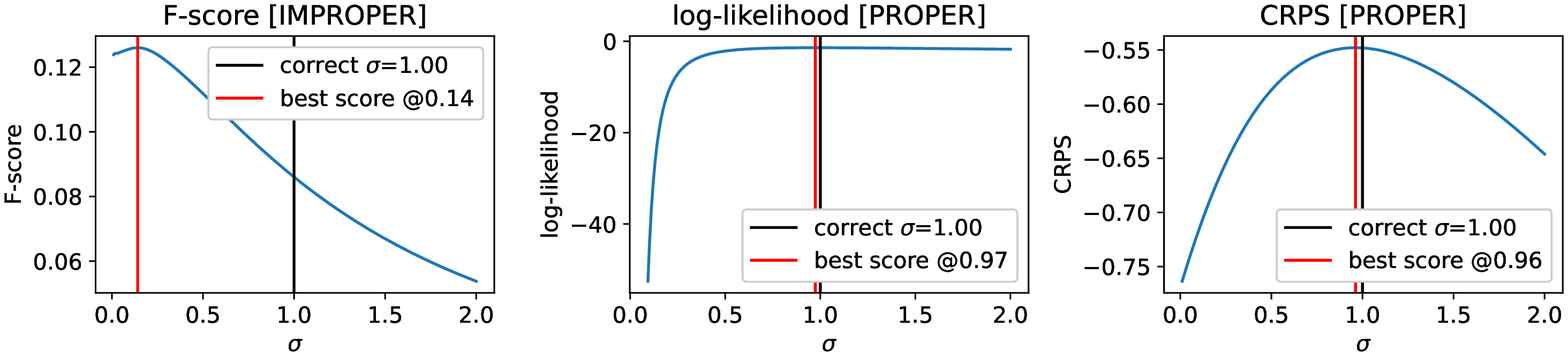}
    \caption{Comparison of the improper F-score with the proper log-likelihood and the CRPS.
    In this example, data is synthetically drawn from a normal distribution with $\mu=0$ and $\sigma=1$.
    Next three different scores are calculated for the estimated distributions with the correct mean but with varying sigmas.
    The scores are plotted against the used sigma values.
    This clearly demonstrates that both the log-likelihood as well as the CRPS reach a maximum close to the true sigma value, while the F-score fails and indicates much smaller sigmas.
    }
    \label{fig:unproper}
\end{figure}

In literature, various scoring functions have been introduced with the aim of measuring predictive performance.
However, the concept of proper scores demands that a score just improves when the predicted distribution approaches the true distribution (compare Sec~\ref{proper}).
As an example of an improper score we will discuss is the so called F-score, introduced to measure of the amount of outliers.
Even though this score can be found in many publications including those related to flagship missions in astronomy, we argue that it is not a proper score and should thus \emph{NOT} be used.
The F-score is defined as the surface area outside of a predefined threshold limit for a set of stacked PDFs.
It can be easily calculated by using the CDF values of the upper and lower threshold limit.
In Fig~\ref{fig:unproper} we demonstrate in a simple example how this score fails.
While both proper scores reach a maximum at the sigma value of the true underlying natural distribution, the F-score indicates that a smaller sigma value should be chosen.
With respect to photometric redshift estimation this means that a simple hedging strategy is available since an estimated PDF will reach better F-scores when the corresponding uncertainty is artificially reduced.
Instead of compelling forecasting models to correctly quantify the true forecast uncertainty, the F-score thus encourages unreasonably small uncertainties.
When artificially decreasing the uncertainty the F-score will improve while the calibration will show a severe underdispersion (see Sec~\ref{pit}).

\section{Deep Convolutional Mixture Density Network (DCMDN)}\label{DCMDN}

\begin{figure*}
\centering
  \includegraphics[trim=0 0 0 0, clip, width=0.95\columnwidth]{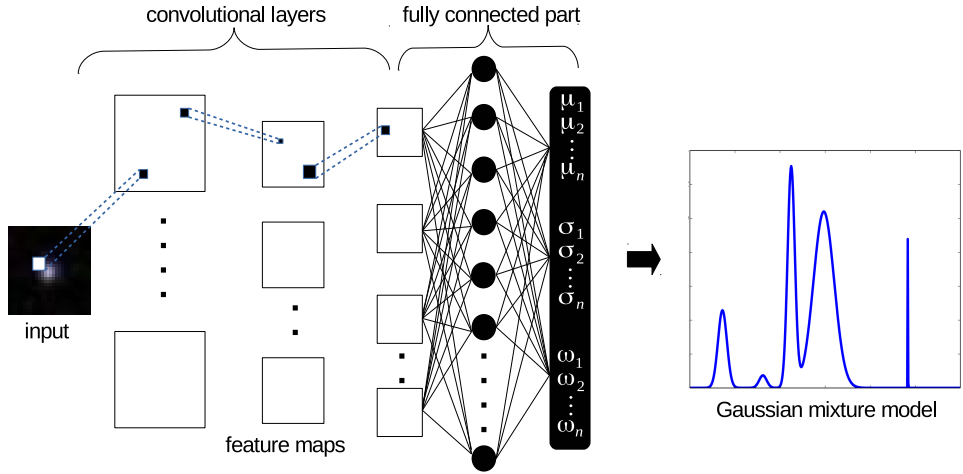}
    \caption{Architecture of the deep convolutional mixture density network, using image matrices as input and generating a vector of parameter of a GMM ($\mathbf{\mu}$, $\mathbf{\sigma}$, $\mathbf{\omega}$).}\label{Fig:dcmdn}
\end{figure*}

In order to estimate photometric redshifts probabilistically based on images as input, we developed a new architecture \citep{DBLP:conf/esann/DIsantoP17}.
The architecture we developed combines a convolutional neural network (CNN) with a mixture density network (MDN) \citep{370fbeadb5584ba9ab2938431fc4f140}.
A CNN is a feed-forward neural network in which several convolutional and sub-sampling layers are coupled with a fully-connected part.
A CNN model has the advantage to deal directly with images, without the need of pre-processing and extracting explicit features, as the network has the ability to capture the important aspects of the inputs autonomously.
This part of the network learns to extract a set of features, the so called feature maps.
The second part utilizes those maps as input for the fully-connected part of the model.
This MDN part of the network generates PDFs as output by providing the parameters necessary to describe a Gaussian mixture model (GMM).
Therefore the architecture is denoted as a DCMDN, whose structure is sketched in Fig~\ref{Fig:dcmdn}.
Instead of training this architecture with the usual log-likelihood loss, we utilized the CRPS.
This allows the model to automatically learn to predict calibrated and sharp PDFs.
Details can be found in \citet{2018A&A...609A.111D}.

\section{Weather Forecast}\label{weather}

Most modern weather forecasts are based on the output of numerical weather prediction (NWP) models which describe the physics of the atmosphere via systems of partial differential equations \citep{BauerEtAl2015}.
Following a profound paradigm shift towards probabilistic forecasting over the past decades, NWP models are usually run multiple times with varying initial conditions or model physics, resulting in a collection of point forecasts, a so called ensemble forecast.
Despite continued advances since the inception of NWP models, ensemble forecasts often exhibit systematic errors that require correction.
This process is referred to as post-processing and aims to remove biases and improve calibration \citep{VannitsemEtAl2021}.

Post-processing models take the ensemble forecasts as input, and produce a probabilistic forecast in the form of a forecast distribution.
A widely used approach is the ensemble model output statistics (EMOS) method proposed by \citet{Gneiting2005}, where the forecast distribution takes the form of a single parametric PDF, the parameters of which are functions of the ensemble predictions.
EMOS models have become standard tools, however, with rapidly growing datasets, inherent limitations prevent them from fully leveraging the complex information in the input data.
For example, ensemble predictions of the variable of interest are often used as sole covariates.
However, usually many more potential predictors (including predictions of variables other than the target quantity) are available, and specifying their functional relations to the forecast distribution parameters is challenging.

To overcome these limitations and make use of modern advances in machine learning, \citet{epub67034} build on the methods developed in \citet{2018A&A...609A.111D} for photometric redshift estimation and propose a neural network (NN) model for post-processing. 
Including location-specific information via embeddings makes the NN models locally adaptive and allows for training a single model jointly for all locations.
This simplifies interpretation and inference, in contrast to standard approaches for post-processing where separate models have to be trained for every location.

In a case study on temperature forecasts at more than 500 weather stations in Germany, the NN models consistently outperform benchmark methods from statistics and machine learning.
The work of \citet{epub67034} has served as basis for various subsequent methodological developments, and the use of neural networks for post-processing has become a central topic of interest in this area \citep{VannitsemEtAl2021}. 

\section{Conclusions}\label{conclusion}

The main challenge of photometric redshifts estimation is the degeneracy of the reconstruction problem, caused by the integration through photometric filters.
For nearly all instrumental filter settings, multimodal redshift distribution have to be expected when just photometric observations are available.
A probabilistic, and multimodal, description of the redshift estimates as a full PDF allows to account for the degeneracy and to propagate the uncertainties correctly.

We introduced proper scores as well as appropriate tools, to evaluate calibration and sharpness of probabilistic forecasts to astronomy already several years ago.
Based on the CRPS, a neural network architecture was trained, to automatically extract calibrated and sharp redshift estimates from plain image data, making the photometric feature extraction step obsolete.
This work lead to an interdisciplinary exchange with researchers from the field of probabilistic weather forecasting.
This is a good example how the exchange of researchers on methodology and machine learning mutually improves scientific work.

We showed how improper evaluation tools and scores that are still commonly used will lead to misleading interpretations.
Research should refrain from utilizing improper methods and adopt proper tools like log-likelihood, CRPS or the PIT for evaluation.
To allow scientist to play with the proposed tools we prepared two examples that can be accessed through:
\textbf{\small
https://slhd.shinyapps.io/shiny/
}
and 

\noindent\textbf{\small
https://colab.research.google.com/drive/15d-IkKiWf0Lhjz\_WxAKkBO5PdmI3AQYl
}

\acknowledgements 
The authors gratefully acknowledge the support of the Klaus Tschira Foundation. Sebastian Lerch further acknowledges support by the Deutsche Forschungsgemeinschaft through SFB/TRR 165 ``Waves to Weather''.





\bibliography{I5-175} 



\appendix



\end{document}